\RequirePackage{fix-cm}
\documentclass[smallextended]{svjour3}       
\smartqed  
\usepackage{soul,color}
\usepackage{graphicx}

\usepackage{verbatim}
\usepackage{caption}
\usepackage{hyperref}
\usepackage{amsmath}
\usepackage{multirow}

\begin{document}

\title{Chest X-ray Image Phase Features for Improved Diagnosis of COVID-19 Using Convolutional Neural Network}


\author{Xiao Qi \and Lloyd Brown \and David J. Foran \and John Nosher \and Ilker Hacihaliloglu
	}


\institute{Xiao Qi\at
           Department of Electrical and Computer Engineering, Rutgers University 
           \email{xq53@scarletmail.rutgers.edu} 
           \and
           Lloyd Brown \at Department of Surgery, Rutgers New Jersey Medical School
           \email{brownl8@njms.rutgers.edu} 
			\and
           David J. Foran \at Rutgers Cancer Institute of New Jersey
           \email{foran@cinj.rutgers.edu} 
           \and
           John Nosher \at Department of Radiology, Rutgers Robert Wood Johnson Medical School
           \email{nosher@rwjms.rutgers.edu} 
           \and
           Ilker Hacihaliloglu \at Department of Biomedical Engineering, Rutgers University \at Department of Radiology, Rutgers Robert Wood Johnson Medical School
           \email{ilker.hac@soe.rutgers.edu} 
}

\date{Received: date / Accepted: date}

\maketitle

\begin{abstract}
\textit{Purpose:}
Recently, the outbreak of the novel Coronavirus disease 2019 (COVID-19) pandemic has seriously endangered human health and life. In fighting against COVID-19, effective diagnosis of infected patient is critical for preventing the spread of diseases. Due to limited availability of test kits, the need for auxiliary diagnostic approach has increased. Recent research has shown radiography of COVID-19 patient, such as CT and X-ray, contains salient information about the COVID-19 virus and could be used as an alternative diagnosis method. Chest X-ray  (CXR) due to its faster imaging time, wide availability, low cost and portability gains much attention and becomes very promising. In order to reduce intra- and inter-observer variability, during radiological assessment, computer aided diagnostic tools have been used in order supplement medical decision making and subsequent management. Computational methods with high accuracy and robustness are required for rapid triaging of patients and aiding radiologist in the interpretation of the collected data.

\noindent\textit{Method:} In this study, we design a novel multi-feature convolutional neural network (CNN) architecture for multi-class improved classification of COVID-19 from CXR images. CXR images are enhanced using a local phase-based image enhancement method. The enhanced images, together with the original CXR data, are used as an input to our proposed CNN architecture. Using ablation studies, we show the effectiveness of the enhanced images in improving the diagnostic accuracy. We provide quantitative evaluation on two datasets and qualitative results for visual inspection. Quantitative evaluation is performed on data consisting of 8,851 normal (healthy), 6,045 pneumonia, and 3,323 Covid-19 CXR scans. 

\noindent\textit{Results:} In Dataset-1, our model achieves 95.57\% average accuracy for a three classes classification, 99\% precision, recall, and F1-scores for COVID-19 cases. For Dataset-2, we have obtained 94.44\% average accuracy, and 95\% precision, recall, and F1-scores for detection of COVID-19.
 
\noindent\textit{Conclusions:} Our proposed multi-feature guided CNN achieves improved results compared to single-feature CNN proving the importance of the local phase-based CXR image enhancement. Future work will involve further evaluation of the proposed method on a larger size COVID-19 dataset as they become available. Training code is available at \url{https://github.com/endiqq/Fus-CNNs_COVID-19}.

\end{abstract}

\section{Introduction}
\label{intro}
Coronavirus disease 2019 (COVID-19) is an infectious disease caused by severe acute respiratory syndrome coronavirus 2 (SARS-CoV-2), a newly discovered coronavirus \cite{singhal2020review,zu2020coronavirus}. In March 2020, the World Health Organization (WHO) declared the COVID-19 outbreak a pandemic. Up to now, more than 9.23 million cases have been reported across 188 countries and territories, resulting in more than 476,000 deaths \cite{dong2020interactive}. Early and accurate screening of infected population and isolation from public is an effective way to prevent and halt spreading of virus. Currently, the gold standard method used for diagnosing COVID-19 is real-time reverse transcription polymerase chain reaction (RT-PCR) \cite{wang2020detection}. The disadvantages of RT-PCR include its complexity and problems associated with its sensitivity, reproducibility, and specificity \cite{bleve2003development}. Moreover, the limited availability of test kits makes it challenging to provide the sufficient diagnosis for every suspected patients in the hyper-endemic regions or countries. Therefore, a faster, reliable and automatic screening technique is urgently required.

In clinical practice, easily accessible imaging, such as chest X-ray (CXR), provides important assistance to clinicians in decision making. Compared to computed tomography (CT) the main advantages of CXR are: Enabling fast screening of patients, being portable, and easy to setup (can be setup in isolation rooms). However, the sensitivity and specificity (radiographic assessment accuracy) of CXR for diagnosing COVID-19 is low compared to CT. This is especially problematic for identifying early stage COVID-19 patients with mild symptoms. This causes larger intra- and inter-observer variability in reading the collected data by radiologists since qualitative indicators can be subtle. Therefore, there is increased demand for computer aided diagnostic method to aid the radiologist during decision making for improved management of COVID-19 disease. 

In view of these advantages and motivated by the need for accurate and automatic interpretation of CXR images, a number of studies based on deep convolutional neural networks (CNNs) have shown quite promising results. Ozturk et al.\cite{ozturk2020automated} proposed a CNN architecture, termed DarkCovidNet, and achieved 87.02\% three class classification accuracy. The method was evaluated on 127 COVID-19, 500 healthy and 500 pneumonia CXR scans. COVID-19 data was obtained from 125 patients. Wang et al.\cite{wang2020covid} built a public dataset named COVIDx, which is comprised of a total of 13975 CXR images from 13870 patient case and developed COVID-Net, a deep learning model. Their dataset had 358 Covid-19 images obtained from 266 patients. Their model achieved 93.3\% overall accuracy in classifying normal, pneumonia, and COVID-19 scans. In \cite{farooq2020covid} a ResNet-50 architecture was utilized to achieve a 96.23\% overall accuracy in classifying four classes, where pneumonia was split into bacterial pneumonia and viral pneumonia. However, there were only eight COVID-19 CXR images used for testing. In \cite{ucar2020covidiagnosis}, 76.37\% overall accuracy was reported on a dataset including 1583 normal, 4290 pneumonia and 76 COVID-19 scans. COVID-19 data was collected from 45 patients. In order to improve the performance of the proposed method, data augmentation was performed on the COVID-19 dataset bringing the total COVID-19 datasize to 1,536. With data augmentation they have improved the overall accuracy 97.2\%. In \cite{siddhartha2020covidlite}, Contrast Limited Adaptive Histogram Equalization (CLAHE) was used to enhance the CXR data. The authors proposed a depth-wise separable convolutional neural network (DSCNN) architecture. Evaluation was performed on 668 normal, 619 pneumonia, and 536 COVID-19 CXR scans. Average reported multi-class accuracy was 96.43\%. Number of patients for the COVID-19 dataset was not available. In \cite{gour2020stacked}, a stacked CNN architecture achieved an average accuracy of 92.74\%. The evaluation dataset had 270 COVID-19 scans from 170 patients, 1139 normal scans from 1015 patients, and 1355 pneumonia scans from 583 patients. In \cite{haghanifar2020covidcxnet}, the reported multi-class average classification accuracy was s 94.2\%. The evaluation dataset included 5000 normal, 4600 pneumonia, and 738 COVID-19 CXR scans. The data was collected from various sources and patient information was not specified. In \cite{apostolopoulos2020covid} transfer learning was investigated for training the CNN architecture. The evaluation dataset included 224 COVID-19, 504 normal, and 700 pneumonia images. 93.48\% average accuracy was reported for three-class classification. The average accuracy increased to 94.72\% if viral pneumonia was included in the evaluation. In \cite{gonzalez2020umls}, performance of three different, previously proposed, CNN architectures was evaluated for multi-class classification. With 2,265 COVID-19 images, the study used the largest COVID-19 dataset reported so far. Average area under the curve (AUC), for classification of COVID-19 from regular pneumonia, was 0.73 \cite{gonzalez2020umls}.

Although numerous studies have shown the capability of CNNs in effective identification of COVID-19 from CXR images, none of these studies investigated local phase CXR image features as multi-feature input to a CNN architecture for improved diagnosis of COVID-19 disease. Furthermore, except \cite{gonzalez2020umls,wang2020covid}, most of the previous work was evaluated on a limited number of COVID-19 CXR scans. In this work we show how local phase CXR features based image enhancement improves the accuracy of CNN architectures for COVID-19 diagnosis. Specifically, we extract three different CXR local phase image features which are combined as a multi-feature image. We design a new CNN architecture for processing multi-feature CXR data. We evaluate our proposed methods on large scale CXR images obtained from healthy subjects as well as subjects who are diagnosed with community acquired pneumonia and COVID-19. Quantitative results show the usefulness of local phase image features for improved diagnosis of COVID-19 disease from CXR scans. 

\section{Material and methods}
\begin{figure}
	\centering
	\includegraphics[width=\linewidth]{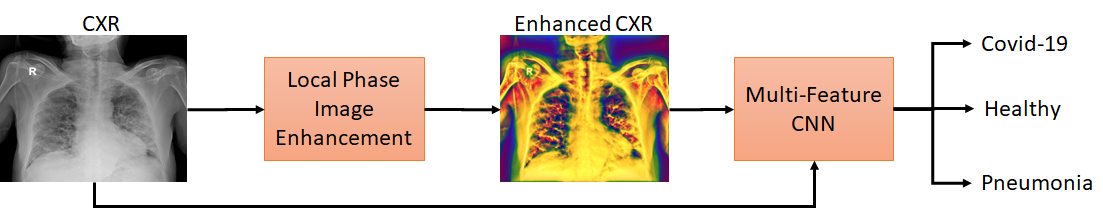}
	\caption{Block diagram of the proposed framework for improved COVID-19 diagnosis from CXR.}
	\label{fig:flowchart} 
\end{figure}
\label{sec:1}
 
Our proposed method is designed for processing CXR images and consists of two main stages as illustrated in Figure \ref{fig:flowchart}: 1- We enhance the CXR images ($CXR(x,y)$) using local phase-based image processing method in order to obtain a multi-feature CXR image ($MF(x,y)$), and 2- we classify $CXR(x,y)$ by designing a deep learning approach where multi feature CXR images ($MF(x,y)$), together with original CXR data ($CXR(x,y)$), is used for improving the classification performance. Next, we  describe how these two major processes are achieved.

\subsection{Image Enhancement}

\noindent In order to enhance the collected CXR images, denoted as $CXR(x,y)$, we use local phase-based image analysis \cite{hacihaliloglu2017localization}. Three different $CXR(x,y)$ image phase features are extracted: 1-  Local weighted mean phase angle ($LwPA(x,y)$), 2- $LwPA(x,y)$ weighted local phase energy ($LPE(x,y)$), and 3- Enhanced local energy attenuation image ($ELEA(x,y)$). $LPE(x,y)$ and $LwPA(x,y)$ image features are extracted using monogenic signal theory where the monogenic signal image ($CXR_{M}$(x,y)) is obtained by combining the bandpass filtered $CXR(x,y)$ image, denoted as $CXR_{B}(x,y)$, with the Riesz filtered components as:

\begin{align*}
CXR_{M}(x,y)= [CXR_{M1},CXR_{M2},CXR_{M3}]\\
=[CXR_{B}(x,y), CXR_{B} \times h_{1}(x,y), CXR_{B}(x,y) \times h_{2}(x,y)]
\end{align*}

\noindent Here $h_{1}$ and $h_{2}$ represent the vector valued odd filter (Riesz filter) \cite{felsberg2001monogenic}. $\alpha$-scale space derivative quadrature filters (ASSD) are used for band-pass filtering due to their superior edge detection \cite{belaid2014alpha}. The $LwPA(x,y)$ image is calculated using:

\begin{equation*}
LwPA(x,y)=arctan(\frac{\sum_{sc}CXR_{M1}(x,y)}{\sqrt{\sum_{sc}CXR_{M1}^{2}(x,y)+\sum_{sc}CXR_{M2}^{2}(x,y)}}).
\end{equation*}

\noindent We do not employ noise compensation during the calculation of the $LwPA(x,y)$ image in order to preserve the important structural details of $CXR(x,y)$. The $LPE(x,y)$ image is obtained by averaging the phase sum of the response vectors over many scales using: 


\begin{equation*}
	\resizebox{\textwidth}{!}{$LPE(x,y)=\{\sum_{sc}|CXR_{M1}(x,y)|-\sqrt{CXR_{M2}^{2}(x,y)+CXR_{M3}^{2}(x,y)}\}\times LwPA(x,y).$}
\end{equation*}

\noindent In the above equation $sc$ represents the number of scales. $LPE(x,y)$ image extracts the underlying tissue characteristics by accumulating the local energy of the image along several filter responses. The $LPE(x,y)$ image is used in order to extract the third local phase image $ELEA(x,y)$. This is achieved by using $LPE(x,y)$ image feature as an input to an L1 norm based contextual regularization method. The image model, denoted as CXR image transmission map ($CXR_{A}(x,y)$), enhances the visibility of lung tissue features inside a local region and assures that the mean intensity of the local region is less than the echogenicity of the lung tissue. The scattering and attenuation effects in the tissue are combined as: $LPE(x,y)=CXR_{A}(x,y)\times ELEA(x,y)+(1-CXR_{A}(x,y))\rho$. Here $\rho$ is a constant value representative of echogenicity in the tissue. In order to calculate $ELEA(x,y)$,  $CXR_{A}(x,y)$ is estimated first by minimizing the following objective function \cite{hacihaliloglu2017localization}:

\begin{equation*}
\frac{\lambda}{2}\parallel CXR_{A}(x,y)-LPE(x,y)\parallel^2_{2} +\sum_{j\in\chi}\parallel W_{j}\circ (D_{j} * CXR_{A}(x,y)) \parallel_{1}.\ 
\end{equation*}	

\begin{figure}
	\centering
	\includegraphics[width=\linewidth]{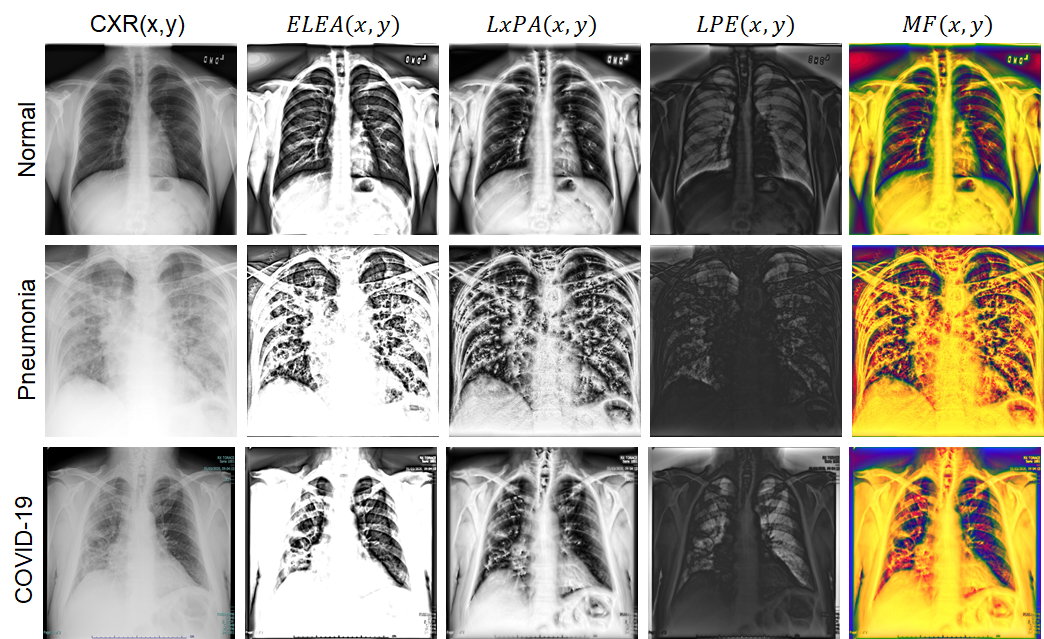}
	\caption{Local phase enhancement of $CXR(x,y)$ images.}
	\label{fig:enhance} 
\end{figure}

\noindent In the above equation $\circ$ represents element-wise multiplication, $\chi$ is an index set, and $*$ is convolution operator. $D_{j}$ is calculated using a bank of high order differential filters \cite{meng2013efficient}. The filter bank enhances the CXR tissue features inside a local region while attenuating the image noise. $W_{j}$ is a weighting matrix calculated using: $W_{j}(x,y)=exp(-\mid D_{j}(x,y) * LPE(x,y)\mid^2)$. In above equation the first part measures the dependence of $CXR_{A}(x,y)$ on $LPE(x,y)$ and the second part models the contextual constraints of $CXR_{A}(x,y)$ \cite{hacihaliloglu2017localization}. These two terms are balanced using a regularization parameter $\lambda$ \cite{hacihaliloglu2017localization}. After estimating $CXR_{A}(x,y)$, $ELEA(x,y)$ image is obtained using: $ELEA(x,y)=[(LPE(x,y)-\rho)/[max(CXR_{A}(x,y),\epsilon)]^\delta]+\rho$. $\delta$ is related to tissue attenuation coefficient $(\eta$) and $\epsilon$ is a small constant used to avoid division by zero \cite{hacihaliloglu2017localization}. Combination of these three types of local phase images as three-channel input creates a new multi-feature image, denoted as $MF(x,y)$. Qualitative results corresponding to the enhanced local phase images are displayed in Figure \ref{fig:enhance}. Investigating Figure \ref{fig:enhance} we can observe that the enhanced local phase images extract new lung features that are not visible in the original $CXR(x,y)$ images. Since local phase image processing is intensity invariant, the enhancement results will not be affected from the intensity variations due to patient characteristics or X-ray machine acquisition settings. The multi-feature image $MF(x,y)$ and the original $CXR(x,y)$ image are used as an input to our proposed deep learning architecture which is explained in the next section.

\subsection{Network Architecture}
\begin{figure}
	\centering
	\includegraphics[width=\linewidth]{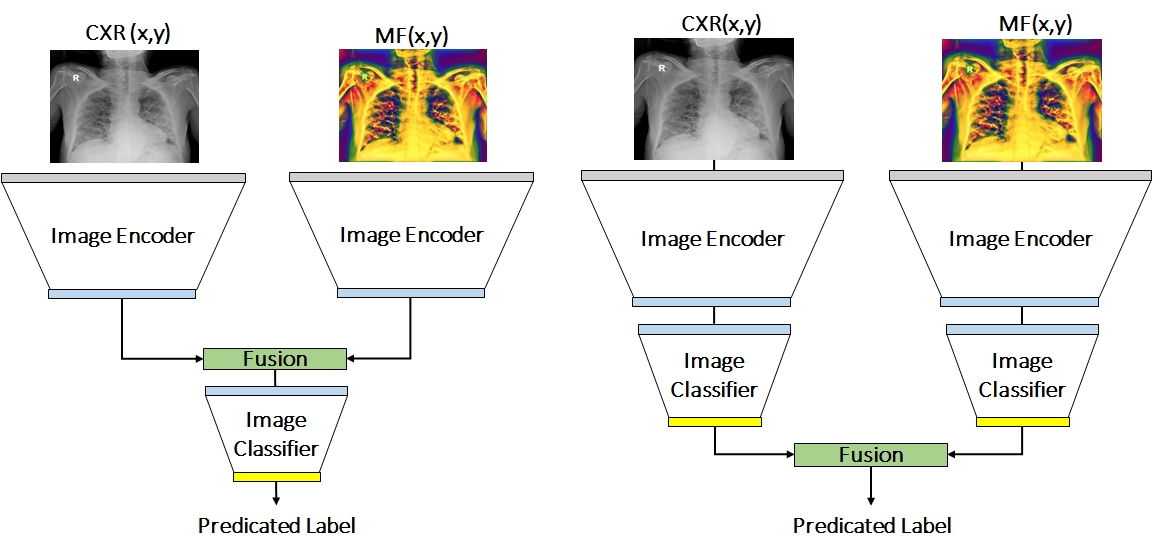}
	\caption{Our proposed multi-feature mid-level (left) and late-level (right) fusion architectures.}
	\label{fig:struct} 
\end{figure}

\noindent Our proposed multi-feature CNN architecture consists of two same convolutional network streams for processing $CXR(x,y)$ images and the corresponding $MF(x,y)$ respectively. Strategies for the optimal fusion of features from multi-modal images is an active area of research. Generally, data is fused earlier when the image features are correlated, and later when they are less correlated \cite{ngiam2011multimodal}. Depending on the dataset, different types of fusion strategies outperform the other \cite{zhou2019review}.  In \cite{alsinan2019automatic}, our group has also investigated early, mid, and late-level fusion operations in the context of bone segmentation from ultrasound data. Late-fusion operation has outperformed the other fusion operations. In \cite{aygun2018multi}, authors have also used late-fusion network, for segmenting brain tumors from MRI data, has outperformed other fusion operations. During this work we design mid-fusion and late-fusion architectures (Fig.\ref{fig:struct}). As part of this work we have also investigate several fusion operations: sum fusion, max fusion, averaging fusion, concatenation fusion, convolution fusion. Based on the performance of the fusion operations and fusion architectures, on a preliminary experiment, we use concatenation fusion operation for both of our architectures. We use the following network architectures as the encoder network: Pretrained AlexNet \cite{krizhevsky2012imagenet}, ResNet50 \cite{he2016deep}, SonoNet64 \cite{baumgartner2017sononet}, XNet(Xception)\cite{chollet2017xception}, InceptionV4(Inception-Resnet-V2)\cite{szegedy2016inceptionv4} and EfficientNetB4 \cite{tan2019efficientnet}. Pretrained AlexNet \cite{krizhevsky2012imagenet} and ResNet50 \cite{he2016deep} have been incorporated into various medical image analysis tasks \cite{litjens2017survey}. SonoNet64 achieved excellent performance in implementation of both classification and localization tasks \cite{baumgartner2017sononet}. XNet(Xception)\cite{chollet2017xception}, InceptionV4 (Inception-Resnet-V2)\cite{szegedy2016inceptionv4} and EfficientNetB4 \cite{tan2019efficientnet} were chosen due to their outstanding performance on recent medical data classification tasks as well as classification of COVID-19 from chest CT data \cite{ha2020identifying,kassani2020automatic}.


\subsection{Dataset}
\label{sec:2}
\noindent We use the following datasets to evaluate the performance of proposed fusion network models: BIMCV \cite{de2020bimcv}, COVIDx \cite{wang2020covid}, and COVID-CXNet \cite{haghanifar2020covidcxnet}. COVID-19 CXR scans from BIMCV\cite{de2020bimcv} and COVIDx \cite{wang2020covid} datasets were combined to generate the 'Evaluation Dataset' (Table \ref{table-eval}). For Normal and Pneumonia datasets we have randomly selected a subset of 2567 images (from 2567 subjects) from the evaluation dataset (Table \ref{table-eval}). In total 2567 images from each class (Normal, Pneumonia, COVID-19) were used during 5-fold cross validation. Table \ref{cross-val} shows the data split for COVID-19 data only. Similar split was also performed for Normal and Pneumonia datasets. In order to provide additional testing for our proposed networks, we have designed a new test dataset which we call 'Test Dataset-2' (Table \ref{table-test2}). The images from Normal and Pneumonia cases which were not included in the 'Evaluation Dataset' were part of the 'Test Dataset-2'. Furthermore, we have included all the COVID-19 scans from COVID-CXNet \cite{haghanifar2020covidcxnet}.

%
%
%

In order to show the improvements achieved using our proposed multi-feature CNN architecture we also trained the same CNN architectures using only $MF(x,y)$ or $CXR(x,y)$ images. We refer to these architectures as mono-feature CNNs. Quantitative performance was evaluated by calculating average accuracy, precision, recall, and F1-scores for each class \cite{ucar2020covidiagnosis,wang2020covid}.

\begin{table}
	\caption{Data distribution of the evaluation dataset}
	\label{table-eval}
	\renewcommand{\arraystretch}{1.2} 
	\centering
	\begin{tabular}{|c|c|c|c|c|c|} 
		\hline
		&\multirow{2}{*}{Normal} & \multirow{2}{*}{Pneumonia} & \multirow{2}{*}{COVID-19\cite{wang2020covid}} & \multirow{2}{*}{COVID-19\cite{de2020bimcv}} & \multirow{2}{*}{COVID-19}\\
		& & & (COVIDx) & (BIMCV) & (Merged) \\ 
		\hline
		\# images   & 8851  & 6045     & 400 & 2167 & 2567\\ 
		\hline
		\# subjects & 8851   & 6031     & 301 & 1183 & 1484\\                            
		\hline
	\end{tabular}
\end{table}

\begin{table}
	\caption{Distribution of 5-fold cross validation dataset split for training, validation, and testing for COVID-19 data only. Same split was also performed for Normal and Pneumonia datasets.}
	\label{cross-val}
	\centering
	\begin{tabular}{|c|c|c|c|c|c|c|} 
		\hline
		&             & k1 & k2 & k3&k4&k5   \\ 
		\hline
		Training data   & \# images   & 1555~  & 1560~     & 1541& 1547& 1529  \\ 
		\hline
		& \# subjects & 890   & 890     & 890 &890 &891   \\ 
		\hline
		Validation data & \# images   & 494~   & 504~      & 512&511&515  \\ 
		\hline
		& \# subjects & 297    & 297       & 297 &297&297  \\ 
		\hline
		Test data       & \# images   & 518    & 503       & 514&509&523   \\ 
		\hline
		& \# subjects & 297    & 297       & 297&297&296   \\                              
		\hline
	\end{tabular}
\end{table}

\begin{table}
	\caption{Data distribution of Test Dataset-2.}
	\label{table-test2}
	\centering
	\begin{tabular}{|c|c|c|c|} 
		\hline
		&Normal & Pneumonia & COVID-19 (COVID-CXNet) \cite{haghanifar2020covidcxnet} \\
		\hline
		\# images   & 6284  & 3478    & 756 \\ 
		\hline
		\# subjects & 6284   & 3464     & Unknown\\                            
		\hline
	\end{tabular}
\end{table}

\section{Results}
The experiments were implemented in Python using Pytorch framework. All models were trained using stochastic gradient descent (SGD) optimizer, cross-entropy loss function, learning rate 0.001 for the first epoch and a learning rate decay of 0.1 every 15 epochs with a mini-batches of size 16. For local phase image enhancement, we have used $sc=2$ and the rest of the ASSD filter parameters were kept same as reported in \cite{hacihaliloglu2017localization}. For calculating $ELEA(x,y)$ images we used $\lambda=2$, $\epsilon=0.0001$, $\eta=0.85$, and $\rho$, the constant related to tissue echogenicity, was chosen as the mean intensity value of $LPE(x,y)$. These values were determined empirically and kept constant during qualitative and quantitative analysis.

\begin{figure}
	\centering
	\includegraphics[width=\linewidth]{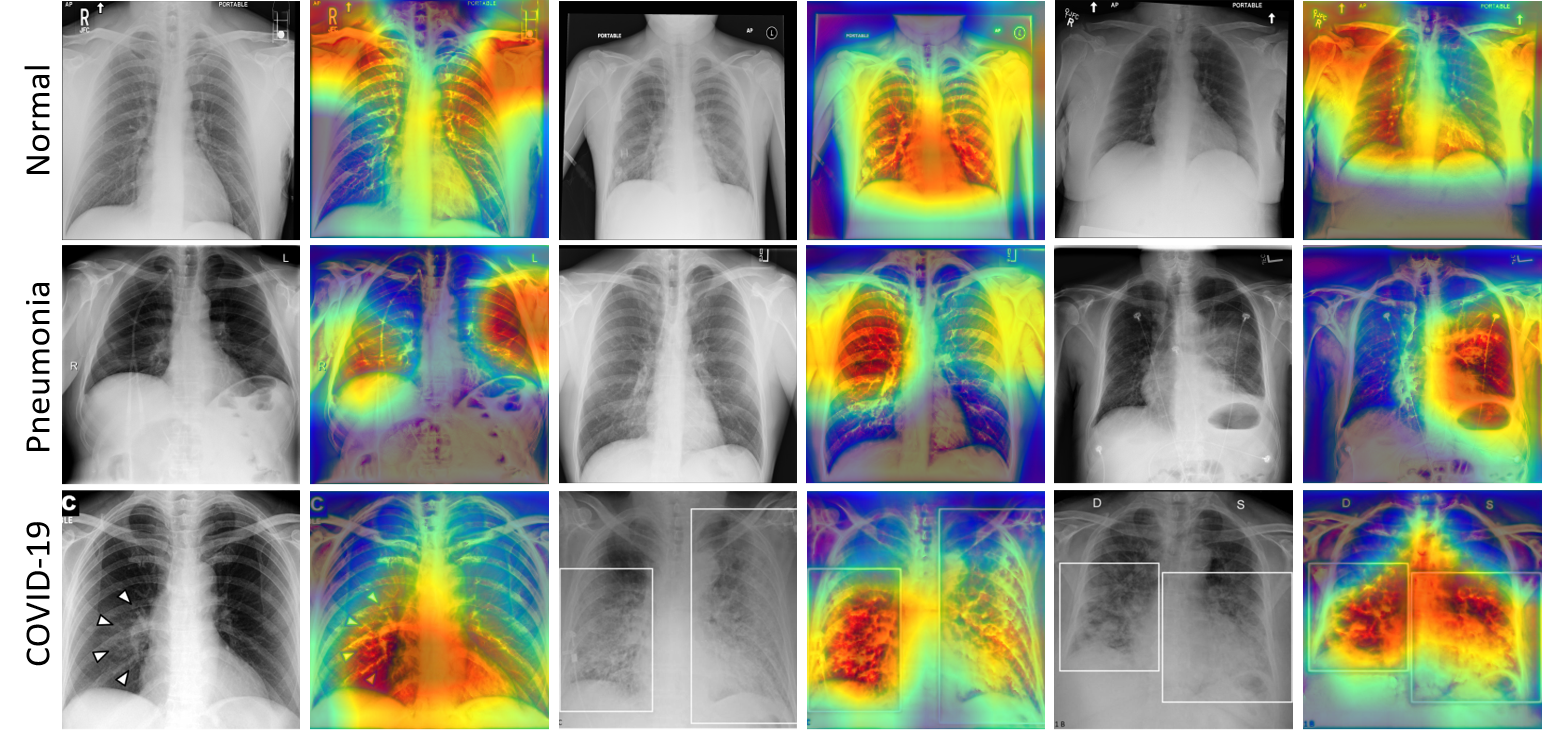}
	\caption{Grad-CAM images \cite{Selvaraju_2019} obtained by late fusion ResNet50 architecture.}
	\label{fig:grad} 
\end{figure}

\paragraph{Qualitative analysis:} 
Gradient-weighted Class Activation Mapping (Grad-CAM) \cite{Selvaraju_2019} visualization of normal, pneumonia, and COVID-19 are presented as qualitative results in Figure \ref{fig:grad}. Investigating Figure \ref{fig:grad} we can see the discriminative regions of interest localized in the normal, pneumonia, and COVID-19 data.  

\begin{figure}
	\centering
	\includegraphics[width=\linewidth]{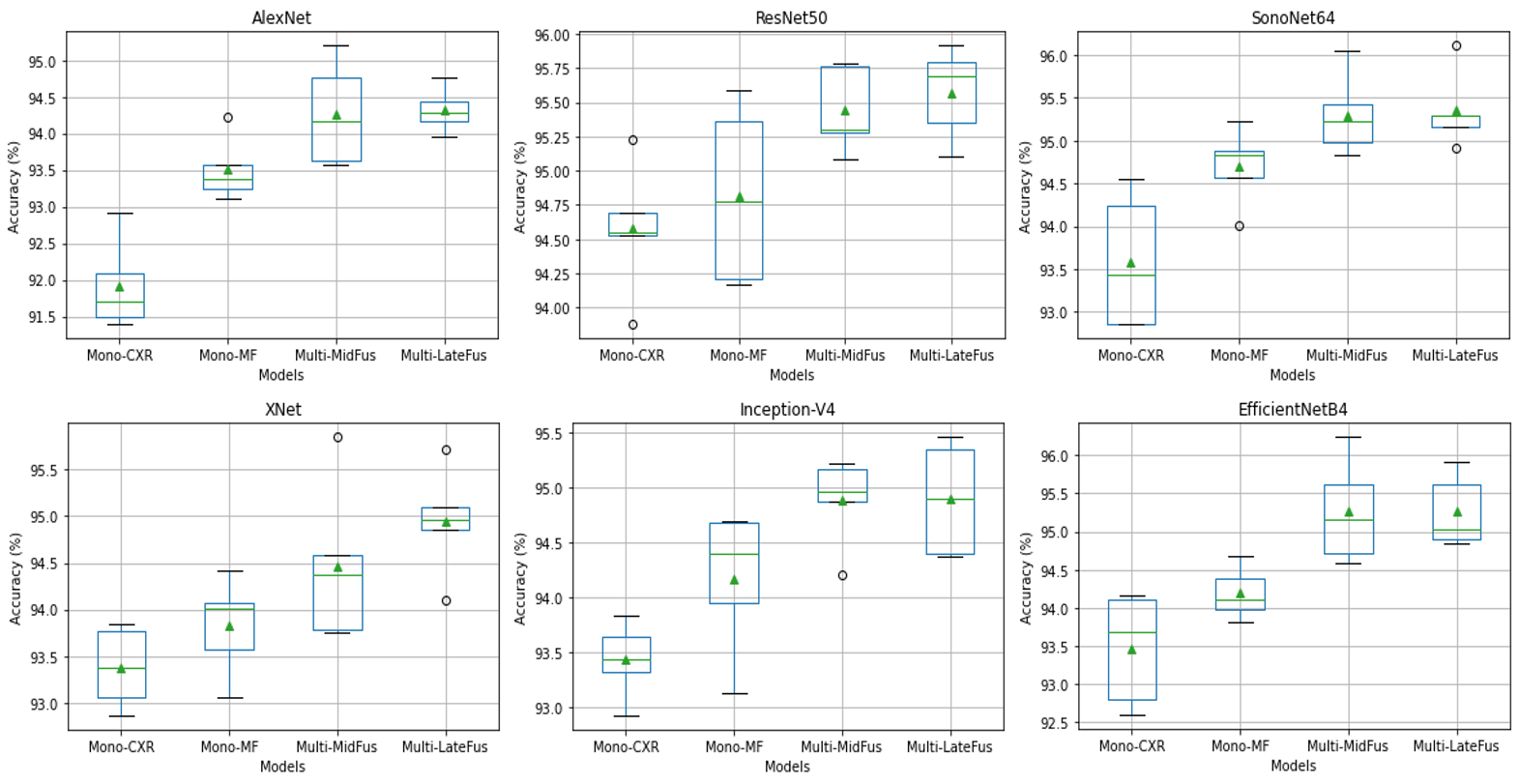}
	\caption{A quantitative results showing classification accuracy of different models for Evaluation Dataset}
	\label{fig:box-dataset1} 
\end{figure}

\begin{figure}
	\centering
	\includegraphics[width=\linewidth]{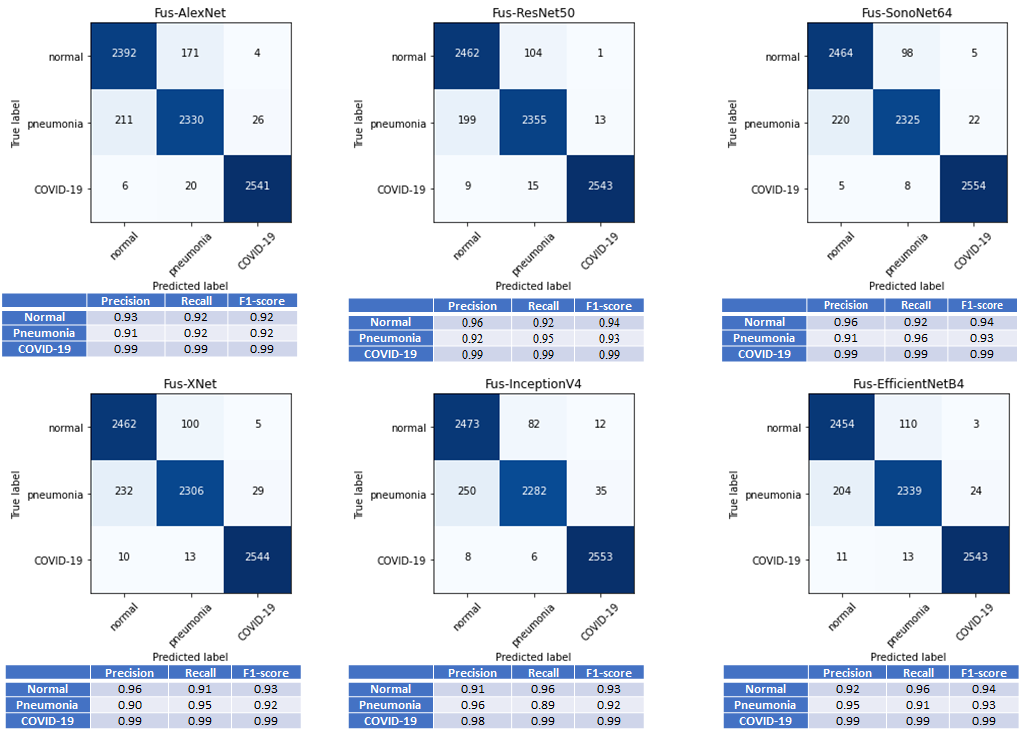}
	\caption{Confusion matrix, and average precision, recall and F1-scores obtained from 5-fold cross validation on 'Evaluation Data' using all multi-feature network models.}
	\label{fig:cm} 
\end{figure}

\paragraph{Quantitative analysis of Evaluation Dataset:} 
Table \ref{tab:eval-data} shows average accuracy of the 5-fold cross validation on the 'Evaluation Dataset' for mono-feature CNN architectures as well as the proposed multi-feature CNN architectures. A Box and Whisker plot is presented in Figure \ref{fig:box-dataset1}.
In most of the investigated network designs $MF(x,y)$-based mono-feature CNN architectures outperform $CXR(x,y)$-based mono-feature CNN architectures. The best average accuracy is obtained when using our proposed multi-feature ResNet50 \cite{he2016deep} architecture. All multi-feature CNNs with mid- and late-fusion operation compared with mono-feature CNNs, with original $CXR(x,y)$ images as input, achieved statistically significant difference in terms of classification accuracy (p$<$0.05 using a paired t-test at $\%5$ significance level). Except SonoNet64 \cite{baumgartner2017sononet}, XNet(Xception)\cite{chollet2017xception}, and InceptionV4(Inception-Resnet-V2)\cite{szegedy2016inceptionv4}, all multi-feature CNNs with mid-fusion operation compared with mono-feature CNNs with $MF(x,y)$ images as input show statistically significant difference in terms of classification accuracy (p$ < $0.05 using a paired t-test at $\%5$ significance level). We did not find any statistical significant difference in the average accuracy results between the middle-level and late-fusion networks (p$>$0.05 using a paired t-test at $\%5$ significance level). Figure \ref{fig:cm} presents confusion matrix results together with average precision, recall, and F1-scores for all multi-feature late-fusion CNN architectures. One important aspect observed from the presented results we can see that almost all the investigated multi-feature networks achieved very high precision, recall, and F1-scores for COVID-19 data indicating very few cases were misclassified as COVID-19 from other infected types.

%

\begin{table}
	\caption{ Mean overall accuracy after 5-fold cross validation on 'Evaluation Data' using mono-feature CNNs and multi-feature CNNs. Bold denotes the best results obtained.}
	\label{tab:eval-data}
	\centering
	\begin{tabular}{|c|c|c|c|} 
		\hline
		& AlexNet & ResNet50 & SonoNet64  \\ 
		\hline
		CXR(x,y)& 91.9$\pm$ 0.55& 94.58$\pm$ 0.43& 93.59$\pm$0.7 \\ 
		\hline
		MF(x,y)& 93.51$\pm$ 0.39 & 94.82$\pm$0.58 &94.70$\pm$0.4	 \\                            
		\hline
		Middle Fusion& 94.27$\pm$0.64 &95.44$\pm$0.28 &95.30$\pm$0.42	  \\                            
		\hline
		Late Fusion& \textbf{94.32$\pm$0.27}	& \textbf{95.57$\pm$0.3} & \textbf{95.35$\pm$0.4} \\
                   
		\hline
		& Xception & InceptionV4 & EfficientNetB4 \\ 
		\hline
		CXR(x,y) & 93.38$\pm$0.38 & 93.43$\pm$0.31 & 93.47$\pm$0.62 \\ 
		\hline
		MF(x,y)	&93.83$\pm$0.47	&94.17$\pm$0.59	&94.19$\pm$0.45 \\                            
		\hline
		Middle Fusion &94.47$\pm$0.76 &94.89$\pm$0.36 &95.26$\pm$0.61  \\                            
		\hline
		Late Fusion	& \textbf{94.95$\pm$0.52}	& \textbf{94.90$\pm$0.46}	& \textbf{95.26$\pm$0.43} \\ 
		\hline

	\end{tabular}
\end{table}

\begin{figure}
	\centering
	\includegraphics[width=\linewidth]{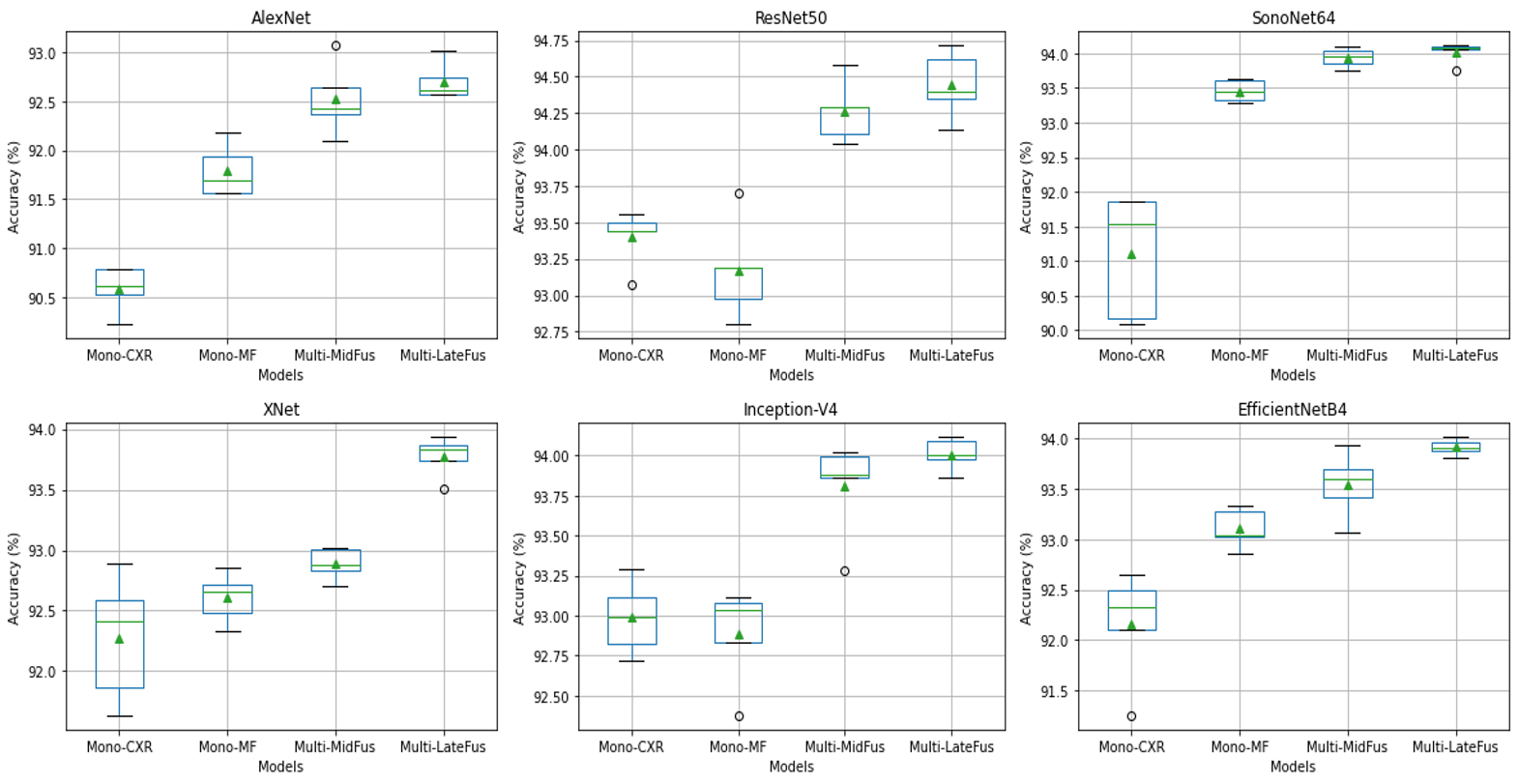}
	\caption{A quantitative results showing classification accuracy of different models for Test Dataset-2}
	\label{fig:box-dataset2} 
\end{figure}

\begin{figure}
	\centering
	\includegraphics[width=\linewidth]{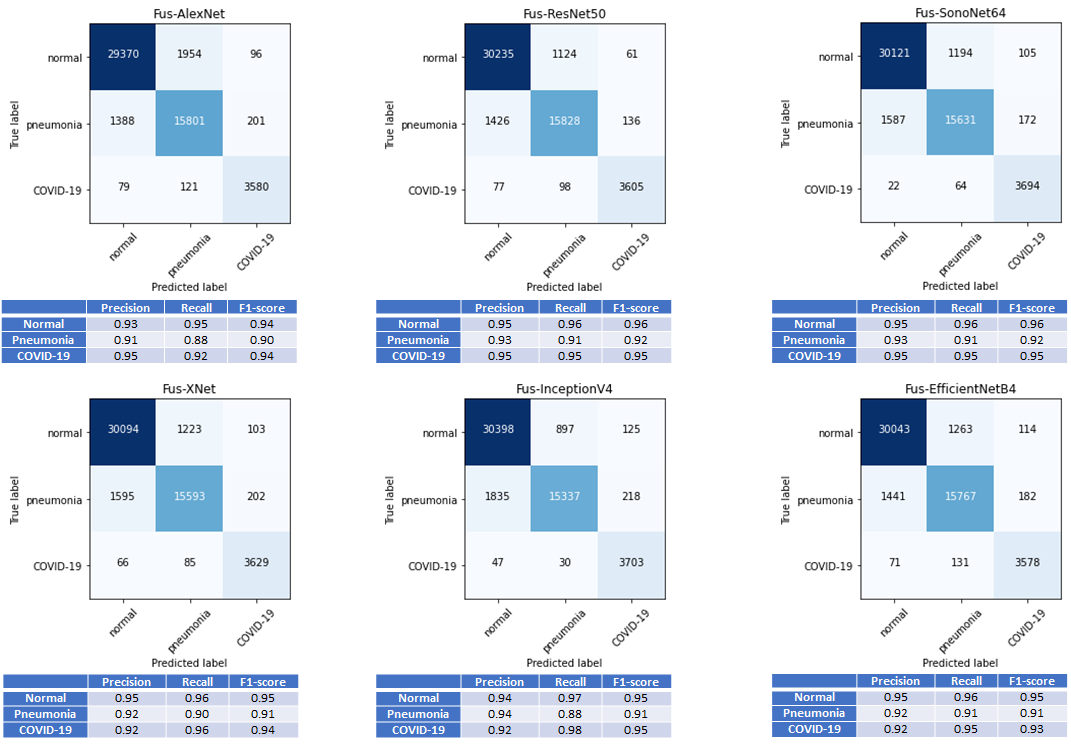}
	\caption{Confusion matrix, and average precision, recall and F1-scores obtained from 5-fold cross validation on 'Test Dataset-2' using all multi-feature network models.}
	\label{fig:cm2} 
\end{figure}

\paragraph{Quantitative analysis of Test Dataset-2:}
Multi-feature ResNet50 provides the highest overall accuracy shown in Table \ref{tab:compare2}, which is consistent with the quantitative result achieved with the 'Evaluation Dataset'. Figure \ref{fig:box-dataset2} shows a Box and Whisker plot for each network. All multi-feature CNNs with late-fusion operation compared with mono-feature CNNs, with original $CXR(x,y)$ images as input, achieved statistically significant difference in terms of classification accuracy (p$<$0.05 using a paired t-test at $\%5$ significance level). Except XNet(Xception)\cite{chollet2017xception}, all the multi-feature CNNs with mid fusion operation compared with mono-feature CNNs with original $CXR(x,y)$ images as input achived statistically significant difference in terms of classification accuracy (p$<$0.05 using a paired t-test at $\%5$ significance level). Except XNet(Xception)\cite{chollet2017xception}, all multi-feature CNNs with mid-fusion operation compared with mono-feature CNNs with $MF(x,y)$ images as input show statistically significant difference in terms of classification accuracy (p$<$0.05 using a paired t-test at $\%5$ significance level). Similar to 'Evaluation Dataset' results, there was no statistically significant difference in the average accuracy results between the middle-level and late-fusion networks (p$>$0.05 using a paired t-test at $\%5$ significance level) except ResNet50\cite{he2016deep}, and XNet(Xception)\cite{chollet2017xception} architectures. Confusion matrix results, together with average precision recall and F1-score values, for all multi-feature late-fusion CNN architectures evaluated are presented in Figure\ref{fig:cm2}. Similar to the results presented for 'Evaluation Dataset', high precision, recall, and F1-score values are obtained for the COVID-19 data.

\begin{table}
	\caption{Mean overall accuracy after 5-fold cross validation on 'Test Dataset-2' using mono-feature CNNs and multi-feature CNNs. Bold denotes the best results obtained.}
	\label{tab:compare2}
	\centering
	\begin{tabular}{|c|c|c|c|} 
		\hline
		& AlexNet & ResNet50 & SonoNet64  \\ 
		\hline
		CXR(x,y)& 90.59$\pm$0.21 & 93.4$\pm$0.17 & 91.1$\pm$0.8 \\ 
		\hline
		MF(x,y)& 91.97$\pm$ 0.24 & 93.17$\pm$0.3 &93.46$\pm$0.15 \\                            
		\hline
		Middle Fusion& 92.52$\pm$0.32 & 94.26$\pm$0.19 &93.94$\pm$0.13	  \\                            
		\hline
		Late Fusion& \textbf{92.72$\pm$0.17} &\textbf{94.44$\pm$0.2}	& \textbf{94.02$\pm$0.14}	 \\
		\hline
		& Xception & InceptionV4 & EfficientNetB4 \\ 
		\hline
		CXR(x,y) & 92.28$\pm$0.46 & 92.99$\pm$0.2 & 92.16$\pm$0.49 \\ 
		\hline
		MF(x,y)	&92.61$\pm$0.19	&92.89$\pm$0.27	&93.1$\pm$0.17 \\                            
		\hline
		Middle Fusion &92.89$\pm$0.12 &93.8$\pm$0.27 &93.54$\pm$0.29  \\                            
		\hline
		Late Fusion	&\textbf{93.77$\pm$0.15}	&\textbf{94.01$\pm$0.09}	&\textbf{93.91$\pm$0.07} \\ 
		\hline
	\end{tabular}
\end{table}

\section{Discussion and Conclusion}

\begin{figure}
	\centering
	\includegraphics[width=\linewidth]{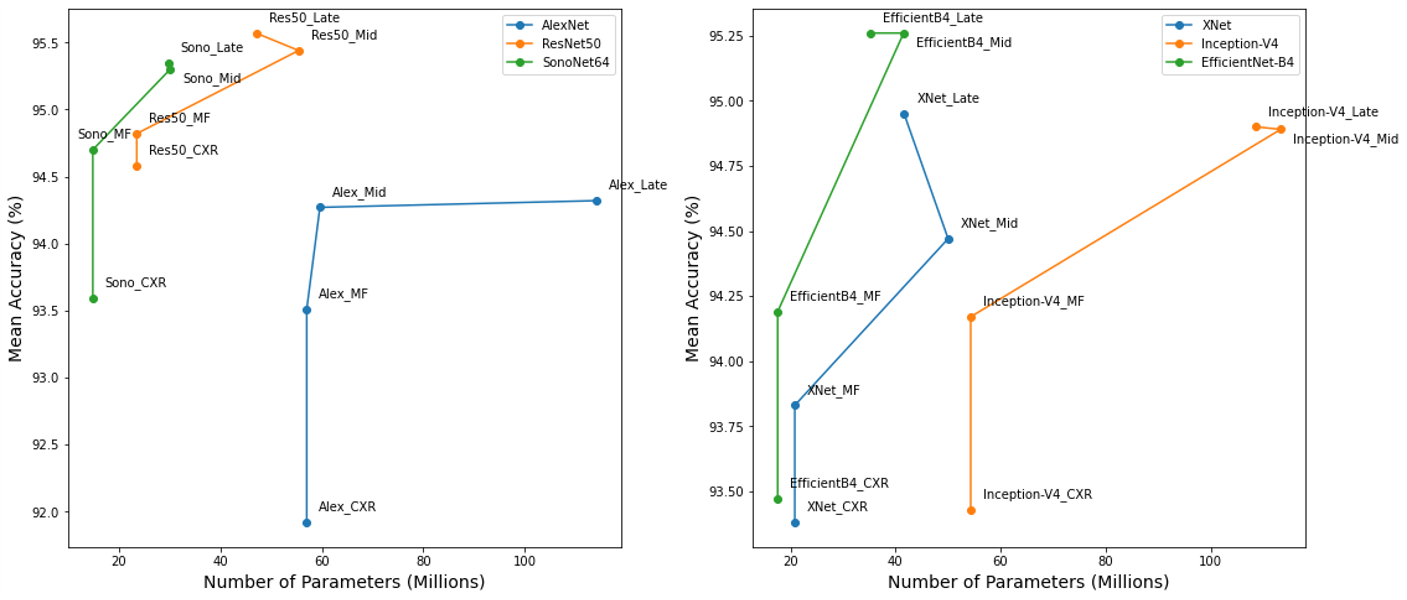}
	\caption{Model Size vs. Overall Accuracy}
	\label{accvsparams} 
\end{figure}

Development of a new computer aided diagnostic methods for robust and accurate diagnosis of COVID-19 disease from CXR scans is important for improved management of this pandemic. In order to provide a solution to this need, in this work, we present a multi-feature deep learning model for classification of CXR images into three classes including COVID-19, pneumonia,and normal healthy subjects. Our work was motivated by the need for enhanced representation of CXR images for achieving improved diagnostic accuracy. To this end we proposed a local phase-based CXR image enhancement method. We have shown that by using the enhanced CXR data, denoted as $MF(x,y)$, in conjunction with the original CXR data, diagnostic accuracy of CNN architectures can be improved. Our proposed multi-feature CNN architectures were trained on a large dataset in terms of the number of COVID-19 CXR scans and have achieved improved classification accuracy across all classes. One of the very encouraging result is the proposed models show high precision, recall, and F1-scores on the COVID-19 class for both testing datasets. In addition, except for AlexNet\cite{krizhevsky2012imagenet}, all multi-feature CNNs with late fusion operation has less number of parameters compared with corresponding multi-feature CNNs with middle fusion operation (Figure \ref{accvsparams}). Since the image classifier of AlexNet \cite{krizhevsky2012imagenet} is consist of three fully connected layers (fc), which store majority of parameters, AlexNet \cite{krizhevsky2012imagenet} with late fusion operation almost double the number of parameters compared with middle fusion operation. The rest of networks have only one or no fc layer in the image classifiers. Finally, compared to previously reported results, our work achieves the highest three class classification accuracy on a significantly larger COVID-19 dataset (Table \ref{tab:literature}). This will ensure few false positive cases for the COVID-19 detected from CXR images and will help alleviate burden on the healthcare system by reducing the amount of CT scans performed. While the obtained results are very promising, more evaluation studies are required specifically for diagnosing early stage COVID-19 from CXR images. Our future work will involve the collection of CXR scans from early stage or asymptotic COVID-19 patients. We will also investigate the design of a CXR-based patient triaging system.

\begin{table}[]
	\begin{center}
		\caption{Comparison of proposed method with recent state-of-the-art methods for COVID-19
			detection using CXR images}
		\label{tab:literature}
		\renewcommand{\arraystretch}{1.2} 
		\begin{tabular}{c|c|l|l|c}
			\hline 
			\textbf{Study} & \textbf{Method} & \multicolumn{2}{c|}{\textbf{Dataset} } &\textbf{Acc (\%)}\\
			\hline
			\multirow{4}{*}{Wang et al. \cite{wang2020detection}} & \multirow{4}{*}{COVID-Net} & \textbf{Training data:} & \textbf{Testing data:} &\multirow{4}{*}{93.3}\\
			& & 7966 Normal & 100 Normal &  \\
			& & 5438 Pneumonia & 100 Pneumonia &  \\
			& & 258 COVID-19 & 100 COVID-19  & \\ \hline
			\multirow{3}{*}{Ozturk et al. \cite{ozturk2020automated}} & \multirow{3}{*}{DarkCovidNet} & \multicolumn{2}{c|}{500 Normal}
			& \multirow{3}{*}{87.02}\\
			& & \multicolumn{2}{c|}{500 Pneumonia} &\\
			& & \multicolumn{2}{c|}{127 COVID-19}  &\\ \hline
			\multirow{4}{*}{Haghanifar et al. \cite{haghanifar2020covidcxnet}} & \multirow{4}{*}{UNet+DenseNet} & \textbf{Training data:} & \textbf{Testing data:} &\multirow{4}{*}{87.21}\\
			& & 3000 Normal & 724 Normal &  \\
			& & 3400 Pneumonia & 672 Pneumonia &  \\
			& & 400 COVID-19 & 144 COVID-19  & \\ \hline
			Siddhartha and  & \multirow{3}{*}{COVIDLite} & \multicolumn{2}{c|}{668 Normal} & \multirow{3}{*}{96.43}\\
			Santra \cite{siddhartha2020covidlite} & & \multicolumn{2}{c|}{619 Viral Pneumonia} & \\
			& & \multicolumn{2}{c|}{536 COVID-19}  & \\ \hline
			\multirow{5}{*}{} & \multirow{5}{*}{VGG19} & \textbf{Testing data 1:} & \textbf{Testing data 2:} &\multirow{5}{*}{}\\
			& & 504 Normal & 504 Normal &  \\
			Apostolopoulos and & & 700 Bacterial & 714 Viral \& &  93.48 \& \\
			Mpesiana \cite{apostolopoulos2020covid} & & Pneumonia & Bacterial Pneumonia&  94.72\\
			& & 224 COVID-19 & 224 COVID-19  & \\ \hline
			\multirow{4}{*}{Proposed Method} & \multirow{4}{*}{Fus-ResNet50} & \textbf{Testing data 1:} & \textbf{Testing data 2:} &\multirow{4}{*}{}\\
			& & 2567 Normal & 6284 Normal & 95.57 \& \\
			& & 2567 Pneumonia & 3478 Pneumonia & 94.44 \\
			& & 2567 COVID-19 & 756 COVID-19  & \\ \hline
		\end{tabular}
	\end{center}
\end{table}

\begin{acknowledgements}
The authors are thankful to all the research groups, and national agencies worldwide who provided the open source X-ray images.
\end{acknowledgements}

%
\section*{Compliance with ethical standards}

\textbf{Funding:} Nothing to declare.

\noindent\textbf{Ethical approval} The article uses open source datasets.

\noindent\textbf{Conflict of interest} The authors declare that they have no conflict of interest.

\bibliographystyle{spmpscicopy}      
\bibliography{manuscript}   

\end{document}